\documentclass[aps,prd,twocolumn,superscriptaddress]{revtex4}
\usepackage{amsmath,amsthm}
\usepackage{aas_macros}
\usepackage{graphicx}
\usepackage[hypertex]{hyperref}
\usepackage{caption} 

\def\beq{\begin{equation}}
\def\eeq{\end{equation}}
\def\bea{\begin{eqnarray}}
\def\eea{\end{eqnarray}}

\begin{document}

\title{Phenomenology of {\it Gravitational Aether}\\ as a solution to the {\it Old} Cosmological Constant Problem}

\author{Siavash Aslanbeigi}\email{saslanbeigi@perimeterinstitute.ca}
\affiliation{Perimeter Institute
for Theoretical Physics, 31 Caroline St. N., Waterloo, ON, N2L 2Y5,Canada}
\affiliation{Department of Physics and Astronomy, University of Waterloo, Waterloo, ON, N2L 3G1, Canada}
\author{Georg Robbers}
\affiliation{Max Planck Institute for Astrophysics, Karl-Schwarzschild-Str. 1, 85741 Garching, Germany}
\author{Brendan Z. Foster}
\affiliation{Foundational Questions Institute, PO Box 3022, New York, NY 10163}
\author{Kazunori Kohri}
\affiliation{Cosmophysics group, Theory Center, IPNS, KEK, and The University for Advanced Study (Sokendai), Tsukuba 305-0801, Japan}
\affiliation{Department of Physics, Tohoku University, Sendai 980-8578, Japan}
\author{Niayesh Afshordi}\email{nafshordi@perimeterinstitute.ca}
\affiliation{Perimeter Institute
for Theoretical Physics, 31 Caroline St. N., Waterloo, ON, N2L 2Y5,Canada}
\affiliation{Department of Physics and Astronomy, University of Waterloo, Waterloo, ON, N2L 3G1, Canada}


\begin{abstract}
One of the deepest and most long-standing mysteries in physics has been the huge discrepancy between the observed vacuum density and our expectations from theories of high energy physics, which has been dubbed the {\it Old} Cosmological Constant problem.  One proposal to address this puzzle at the semi-classical level is to decouple
quantum vacuum from space-time geometry via a modification of gravity that includes an incompressible fluid, known as {\it Gravitational Aether}. In this paper,  we discuss classical predictions of this theory along with its compatibility with cosmological
and experimental tests of gravity. We argue that deviations from General Relativity (GR) in this theory are sourced by pressure or vorticity. In particular, the theory predicts that the gravitational constant for radiation is 33\% larger than that of non-relativistic
matter, which is preferred by (most) cosmic microwave background (CMB), Ly-$\alpha$ forest, and ${}^{7}\text{Li}$ primordial abundance observations, while being consistent with other cosmological tests at $\sim 2\sigma$ level.
It is further shown that all Parametrized Post-Newtonian (PPN) parameters have the standard GR values aside from the anomalous coupling to pressure $\zeta_4$, which has not been directly measured. A more subtle prediction of this model (assuming irrotational aether) is that the (intrinsic) gravitomagnetic effect  is 33\% larger than GR prediction. This is consistent with current limits from LAGEOS and Gravity Probe B at $\sim 2\sigma$ level.
\end{abstract}

\maketitle
\section{Introduction}
The discovery of recent acceleration of cosmic expansion
was one of the most surprising findings in modern
cosmology \cite{Riess:1998cb,Perlmutter:1998np}.
The standard cosmological model (also
known as the concordance model) drives this expansion
with a cosmological constant (CC). While the CC is consistent
with (nearly) all current cosmological observations, it requires
an extreme fine-tuning of more than 60 orders of magnitude,
known as the cosmological constant
problem \cite{RevModPhys.61.1}. More precisely, a covariant regularization of
the vacuum state energy of a Quantum Field Theory(QFT), if it exists,
acts just like the CC in linear order, but has a value many orders
of magnitude larger than what is inferred from observations.

If the QFT prediction of the cosmological constant
is considered reasonable (and in lieu of an extreme fine-tuning), there is no choice but to abandon the
idea that vacuum energy should gravitate. This, however, requires
modifying Einstein's theory of gravity, in which all sources of
energy gravitate. Attempts in this direction have been proposed in the context of massive gravity \cite{Dvali:2007kt}, or braneworld models of extra dimensions such as cascading gravity \cite{deRham:2007xp,deRham:2007rw}, or supersymmetric large extra dimensions ({\it e.g.,} \cite{Burgess:2005cg}). However, efforts to find explicit cosmological solutions that de-gravitate vacuum have proved difficult ({\it e.g.,} \cite{Agarwal:2009gy,deRham:2010tw}).

In \cite{Afshordi:2008xu}, one of us proposed a
novel approach to modified gravity in which the QFT vacuum quantum
fluctuations (of linear order in the metric) are decoupled from gravity through
the introduction of an incompressible perfect fluid called the
Gravitational Aether. In this model, the right hand side of the Einstein field
equation is modified as:
\beq
(8\pi G')^{-1}G_{\mu\nu}=T_{\mu\nu}-\frac{1}{4}T^{\alpha}_{\alpha}g_{\mu\nu}+T'_{\mu\nu}
\label{1}
\eeq
\beq
T'_{\mu\nu}=p'(u'_{\mu}u'_{\nu}+g_{\mu\nu}),
\label{2}
\eeq
where $G'$ is the (only) constant of the theory and $T'_{\mu\nu}$ is the
aether fluid which has pressure  $p'$ and four-velocity $u'$. Our metric signature is $(-,+,+,+)$.  Aether
is constrained by requiring the conservation of the energy-momentum
tensor $T_{\mu\nu}$, and the Bianchi identity:
\beq
\nabla^{\mu}T'_{\mu\nu}=\frac{1}{4}\nabla_{\nu}T,
\label{3}
\eeq
which can be written in a similar form to the relativistic hydrodynamic equations:
\bea
p' {\bf \nabla\cdot u'} = -\frac{1}{4}\dot{T}, \label{aether_cont}\\
p'\dot{\bf u}'=-{\bf \nabla_{\perp}}\left(p'-\frac{T}{4}\right), \label{aether_euler}
\eea
 where $\dot{} \equiv {\bf u'\cdot \nabla}$, and $\nabla_{\perp}$ is the gradient normal to ${\bf u}'$ four-vector. Notice that Eqs. (\ref{aether_cont}-\ref{aether_euler}) can be combined to find a parabolic equation for the evolution of ${\bf u'}$, which generically has a well-defined initial value problem, at least for a finite time \footnote{See Appendix \ref{app2} for an example of explicit solutions at linearized level. While aether singularities may develop in the vicinity (or inside the horizon) of black holes, as we demonstrate throughout the paper, solutions exist  for all other situations of physical relevance.}.

This modification of Einstein equations (\ref{1}-\ref{2}), if self-consistent and in
agreement with other experimental bounds on gravity,
could potentially constitute a solution to the old cosmological
constant problem. We will show in this paper that none of
these experimental bounds, as of yet, rule out this theory
(at $\sim2\sigma$ level)
and
that it is surprisingly similar to general relativity \footnote{However, we should note that there is no known action principle that could lead to Eqs. (\ref{1}-\ref{3}).}.

Nevertheless, the new cosmological constant problem, {\em i.e.} the present-day acceleration of cosmic expansion is not addressed by the original gravitational aether proposal. In \cite{PrescodWeinstein:2009mp,Afshordi:2010eq}, it is argued that quantum gravity effects in the presence of astrophysical black holes can naturally explain this phenomenon. In this proposal, the formation of black holes leads to a negative aether pressure, that is set by the horizon temperature of the black holes. However, in the present work we only focus on phenomenological implications of the {\em classical} gravitational aether scenario, and defer study of black hole-dark energy connection, which could be potentially very important on cosmological scales at late times.  Instead, we use a standard cosmological constant to model the late-time acceleration of cosmic expansion. Throughout the paper we set the speed of light c=1.

\section{Cosmological Constraints on Gravitational Aether}\label{cosmology}
If the energy-momentum tensor of matter, $T_{\mu\nu}$, can be approximated by a perfect fluid with constant equation of
state $p=w\rho$ and four-velocity $u_{\mu}$, direct substitution into Eq. (\ref{1}) shows that if: $u'_{\mu}=u_{\mu}$,$p'=\frac{(1+w)(3w-1)}{4}\rho$,
then the solutions to the gravitational aether theory are identical to those of General Relativity (GR) with a renormalized gravitational constant:
\beq
G_N\rightarrow G_{\rm eff}=(1+w)G_N,
\label{4}
\eeq where $G_{N}=3G'/4$. In other words, the gravitational coupling is not a constant anymore, and can change significantly for fluids with relativistic pressure. Not surprisingly, for vacuum equation of state $w=-1$, $G_{\rm eff} =0$, which implies that vacuum does not gravitate.

In particular, in the case of homogeneous FLRW cosmology where the perfect fluid approximation is valid,
this theory predicts that the effective G that relates geometry to
the matter density $\rho$ in Friedmann equation is different in the matter and radiation eras:
\beq
\frac{G_N}{G_R}\equiv\frac{G_{\rm eff}(w=0)}{G_{\rm eff}(w=1/3)}=\frac{3}{4}.
\label{5}
\eeq
This is the first cosmological prediction of this theory: radiation energy gravitates more
strongly than non-relativistic matter. The expansion history in the radiation
era depends on the product $G\rho_{\rm rad}$, and is constrained through different
observational probes. The constraints are often described as the bound
on the effective number of neutrinos $N_{\nu, {\rm eff}}$, which quantifies the total
radiation density $\rho_{\rm rad}$. However, assuming only photons (that are constrained by CMB observation) and three neutrino species, with no more light particles left over from the very early universe, we can translate the constraints to those on $G_{\rm eff}$ by requiring
$G_{\rm eff}\rho_{\rm rad} (N_{\nu}=3)=G_N\rho_{\rm rad}(N_{\nu}=3+\Delta N_{\nu})$.
In particular, based on standard thermal decoupling of neutrinos, Eq. (\ref{5}) can be translated to $\Delta N_{\nu} =2.5$, at least for a homogeneous universe \footnote{Requiring $G_{\rm eff}\rho_{\rm rad} (N_{\nu}=3)=G_N\rho_{\rm rad}(N_{\nu}=3+\Delta N_{\nu})$, we can determine $\Delta N_{\nu}$ in terms of
$\frac{G_R}{G_N}=\frac{4}{3}$ by using $\rho_{rad}=\frac{\pi^2}{30}g_{*}T_{rad}^4$ where $g_{*}\approx 2+0.45N_{\nu}$. Solving for $\Delta N_{\nu}$ gives $\Delta N_{\nu}\approx 2.5$.}. Using this correspondence, we can now discuss cosmological constraints on the gravitational aether scenario.

\begin{figure}[h!]
\centering
\includegraphics[width=0.5\textwidth]{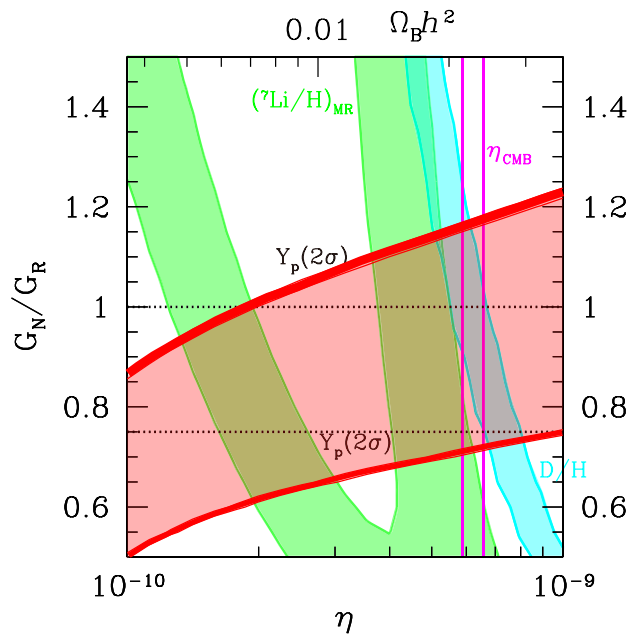}
\caption{Allowed regions with 2 $\sigma$ lines for D/H, $Y_p$ and
${}^{7}\text{Li}/$H are shown. The upper and lower horizontal dashed
lines indicate GR and gravitational aether predictions,
respectively. The thickness of $Y_p$ means the uncertainty in
measurements of neutron lifetime~\cite{Nakamura:2010zzi,Serebrov:2004zf}. We can translate the
vertical axis into $\Delta N_{\nu}$ by using a relation $G_{N}/G_{R}\simeq 1/( 1 + 0.135 \Delta N_{\nu} )$.}
\label{fig1}
\end{figure}

\subsection{Big Bang Nucleosynthesis}

It has been known that the increase of the gravitational constant at
around $T= {\cal O}(1)$ MeV epoch induces earlier freezeout of the
neutron to proton ratio because of a speed-up effect of the increased
cosmic expansion. This raises the abundance of $^{4}$He sensitively
and deuterium (D) mildly, and can lower the abundance of $^{7}$Be
through $^{7}$Be $(n,p)$$^{7}$Li$(p,\alpha)$$^{4}$He (Note that the
second $p$ is  thermal proton). For a relatively large baryon to
photon ratio $\eta \gtrsim 3 \times 10^{-10}$, the dominant mode to
produce $^{7}$Li is the electron capture of $^{7}$Be at a later epoch
through $^{7}$Be + $e^{-}$ $\to$ $^{7}$Li + $\nu_{e}$. Therefore, the
decrease of $^{7}$Be makes the fitting better because so far any
observational $^{7}$Li abundances have been so low that they could not
have agreed with theoretical prediction in Standard BBN (SBBN) at
better than 3 $\sigma$~\cite{Cyburt:2008kw}.

In this study, we adopt the following observational light element
abundances as primordial values: the mass fraction of $^{4}$He,
Y$_{p}= 0.2561 \pm 0.0108 ~({\rm stat})$~\cite{Aver:2010wq},
the deuterium to hydrogen ratio, D/H=$(2.80
\pm 0.20) \times 10^{-5}$~\cite{Pettini:2008mq},  and the  $^7$Li to
hydrogen ratio ${\rm Log}_{10}(^7\text{Li}/\text{H}) = -9.63 \pm
0.06$~\cite{Melendez:2004ni} \footnote{See also ${\rm
Log}_{10}(^7\text{Li}/\text{H}) = -9.90 \pm
0.09$~\cite{Bonifacio:2006au} for the lower value which makes fitting
worse.}. Theoretical errors come from experimental uncertainties in
cross sections~\cite{Smith:1992yy,Cyburt:2001pp,Cyburt:2008kw} and
neutron lifetime~\cite{Nakamura:2010zzi,Serebrov:2004zf}.

Comparing theoretical prediction with the observational light element
aubndances provides a constraint on $G_N/G_R$. Fig.~\ref{fig1} shows
the results of a comprehensive analysis for ${}^{4}\text{He}$, D, and
${}^{7}\text{Li}$.  We also plotted a band for baryon to photon ratio,
$\eta$ which was reported from CMB observations by WMAP 7-year,
$\eta=(6.225 \pm 0.170)\times 10^{-10}$ in case of
$G_N/G_R=1$~\cite{Komatsu:2010fb}.  Then we can see that every light
element agrees with the Gravitational Aether theory within 2
$\sigma$. It is notable that $^7$Li in this theory fits the data
better than that in SBBN. Performing $\chi^2$ fitting for three
elements with three degree of freedom, however, the model is allowed
only at 99.7$\%$ (3 $\sigma$) in total.

However, notice that the main discrepancy is with deuterium abundance observed in quasar absorption lines, which suffer from an unexplained scatter. Moreover, deuterium could be depleted by absorption onto dust grains that would make its primordial value closer to our prediction (see \cite{Pospelov:2010hj} for a discussion).
\subsection{Power Spectrum of Cosmological Fluctuations}
\label{CMB}

\begin{figure*}
\centering
\includegraphics[bb=0 0 595.3  841.9, angle=-90, width=0.3\textwidth]{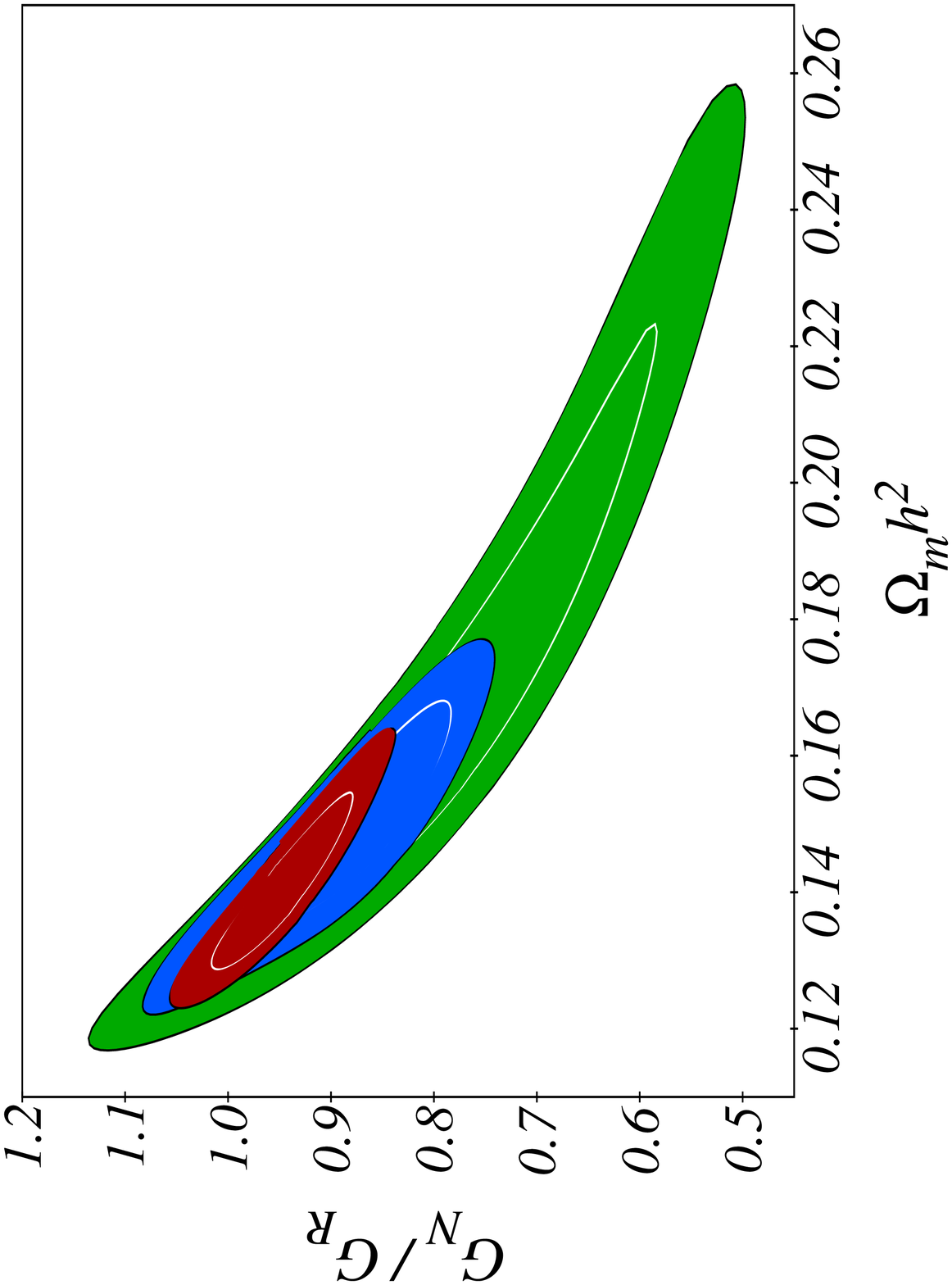}
\includegraphics[bb=0 0 595.3  841.9, angle=-90, width=0.3\textwidth]{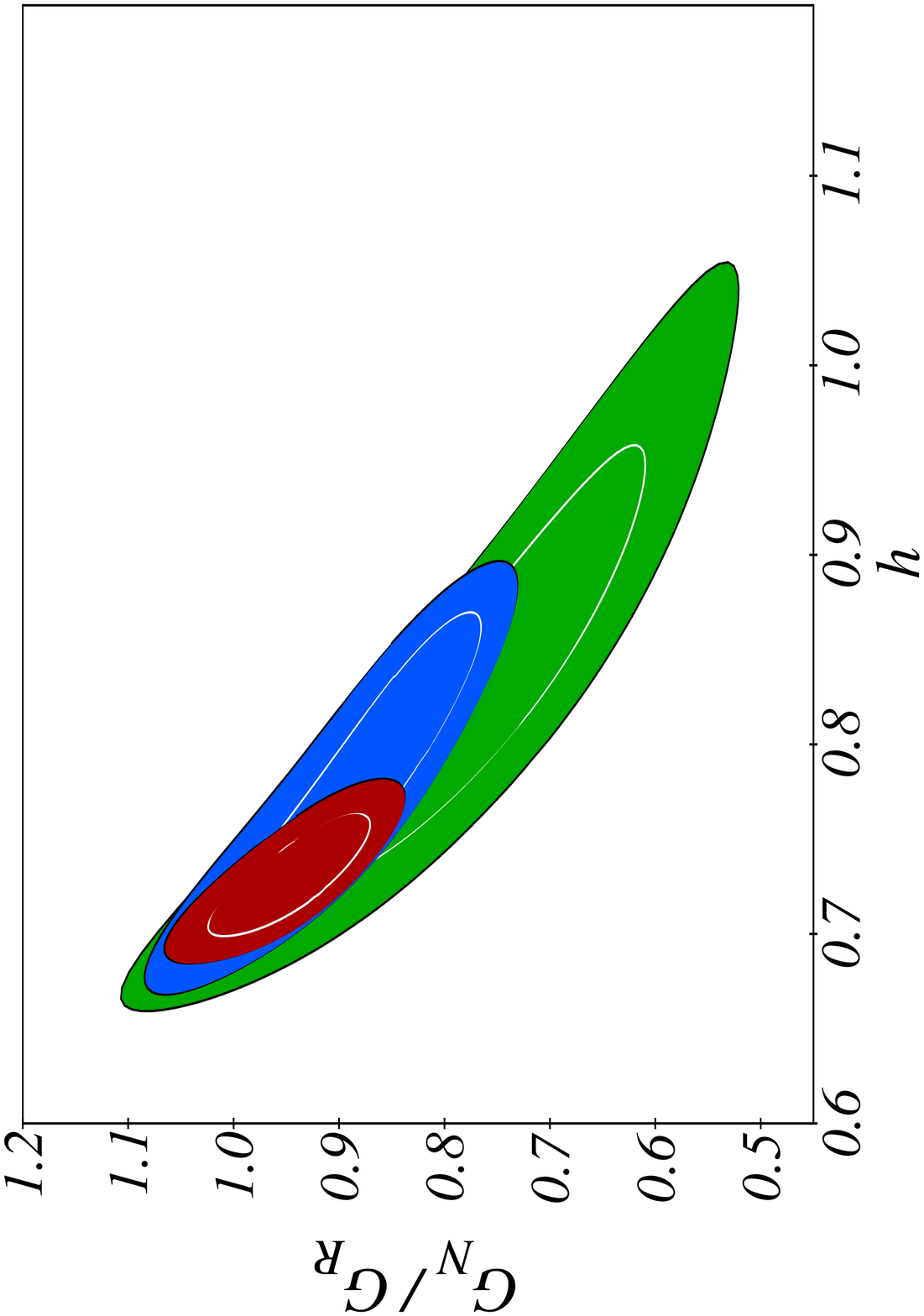}
\includegraphics[bb=0 0 595.3  841.9, angle=-90, width=0.3\textwidth]{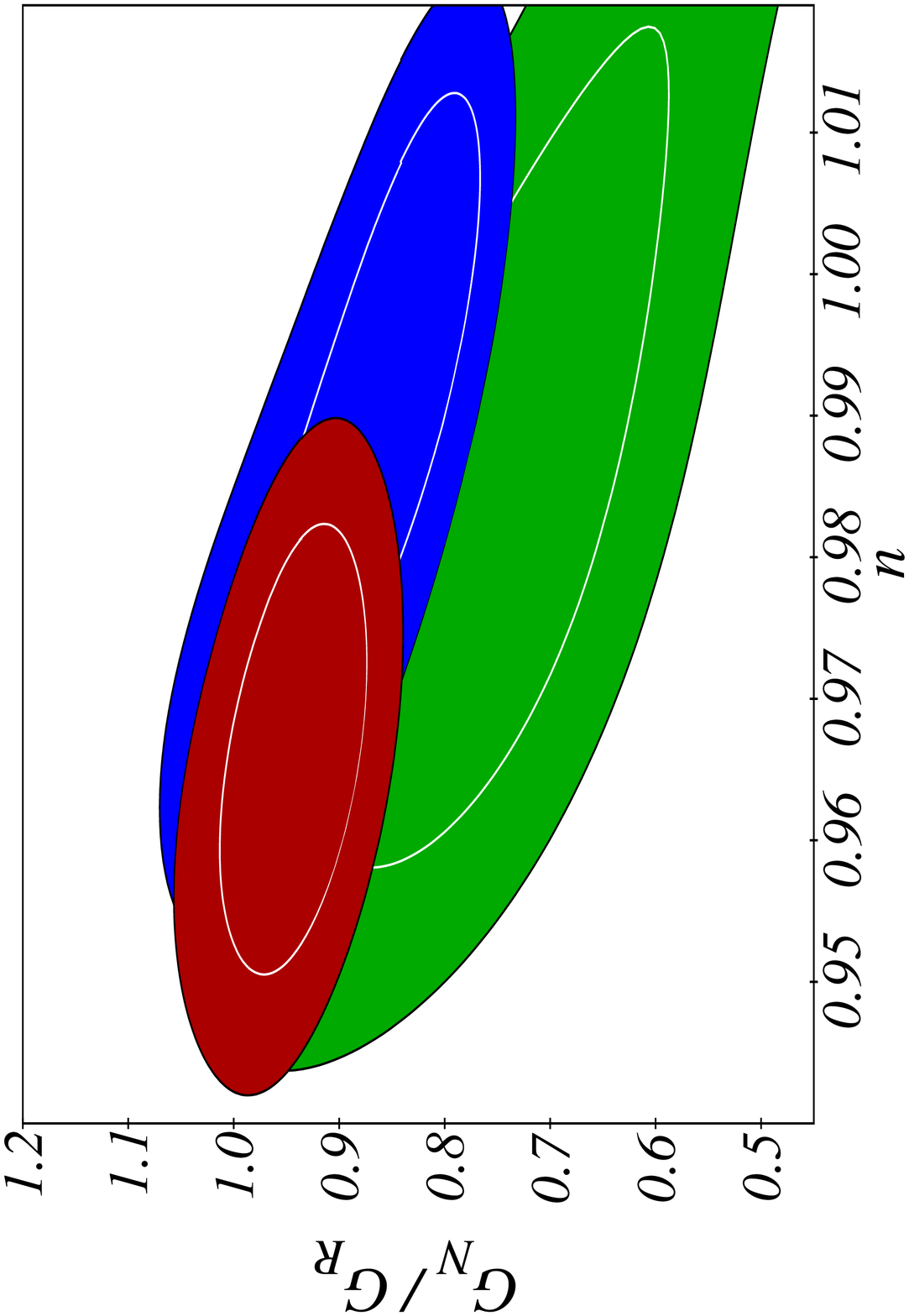}

\captionsetup{margin=10pt,font=small,labelfont=bf}
\caption{Constraints at the $95\%$ confidence level for $G_N/G_R$ from WMAP 7-year (background, green), WMAP+ACT+SPT (middle, blue) and WMAP+ACT+SPT+Sne+BAO+Hubble data (front, red).
         The white lines show the $68\%$ confidence levels. Note that the Gravitational Aether prediction is
	$G_N/G_R=0.75$, while in General Relativity $G_R=G_N$.
        }
\label{fig2a}
\end{figure*}

The Gravitational Aether theory can also be tested by considering the power spectrum
of the CMB, just as a number of publications have recently investigated the
apparent preference for extra relativistic degrees of freedom (see {\it e.g.,} \cite{Hamann:2010bk,Hou:2011ec,Calabrese:2011hg}).
Using a modified version of Cmbeasy~\cite{Doran:2003sy,Doran:2003ua},
 we compute constraints on $G_N/G_R$ from
scalar perturbations in a scenario with three massless neutrino species (details are discussed in Appendix \ref{app2}).
The 7-year CMB data from WMAP~\cite{Komatsu:2010fb} together with small-scale observations from
the Atacama Cosmology Telescope (ACT) ~\cite{Dunkley:2010ge} yield $G_N/G_R=0.73^{+0.31}_{-0.21}$ at $95\%$-confidence.
Just like any additional relativistic component can be compensated by a higher fraction of dark matter in order to keep the time of
matter-radiation equality constant, there is a high amount of degeneracy between $G_N/G_R$ and $\Omega_m h^2$ and $h$ (see Fig.~\ref{fig2a}).
Recent data from the South Pole Telescope (SPT), which measured the CMB power spectrum in the multipole range $650 <\ell<3000$ significantly tightens
the constraint and yields $0.88^{+0.17}_{-0.13}$ (for the combination of ACT and SPT data we have adopted the SPT treatment of foreground nuisance parameters).
A similar effect can be seen when adding Baryonic Acoustic Oscillations (BAO) ~\cite{Percival:2009xn} and constraints on the Hubble rate.
Here we adopt the value of $H_0=73.8\pm2.4\ \mathrm{km^{-1}\ Mpc^{-1}}$~\cite{Riess:2011yx}. Then, by breaking the
degeneracy between the matter content and $h$, the combination WMAP+ACT+BAO+Sne+Hubble result in $G_N/G_R=0.89^{+0.13}_{-0.11}$.
The supernovae data of the Union catalog~\cite{Amanullah:2010vv} do not
significantly contribute to this constraint.
We note that both cases, i.e. adding either SPT data or adding the Hubble constraints to the basic WMAP+ACT set, move the gravitational Aether
value of $G_N/G_R=0.75$ to the border or just outside of the $95\%$ confidence interval,
while the General Relativity value of $G_N/G_R=1.0$ is well compatible with all combinations of data. Consequently, the full combination
of data, i.e. WMAP+ACT+SPT+Hubble+BAO+Sne, constrains $G_N/G_R$ to $0.94^{+0.10}_{-0.09}$.

In contrast, observational constraints at lower redshifts, in particular  data of the Ly-$\alpha$ forest \cite{Seljak:2006bg} prefer the Aether prediction.
Furthermore, additional degeneracies with e.g. the Helium mass fraction $Y_p$ might shift the preferred values.
Combining WMAP+ACT+Sne with  the Ly-$\alpha$ forest constraints  yields,
$G_N/G_R = 0.68^{+0.32}_{-0.25}$ at 95\% level, with  $Y_p$ as a free parameter.
However, we should note that this result is more prone to systematic uncertainties due to the quasilinear nature of the Ly-$\alpha$ forest.
Also, including the SPT data in this combination changes this result to the higher value of $0.90^{+0.27}_{-0.23} $.
 A summary of the constraints with different combinations of data is provided in Table~\ref{tab-constraints}.

Future CMB observations by the Planck satellite, as well as ground-based observatories
are expected to improve this constraint dramatically over the next five years,
and thus confirm or rule out this prediction.

\begin{figure*}
\centering
\includegraphics[bb=0 0 595.3  841.9, angle=-90, width=0.3\textwidth]{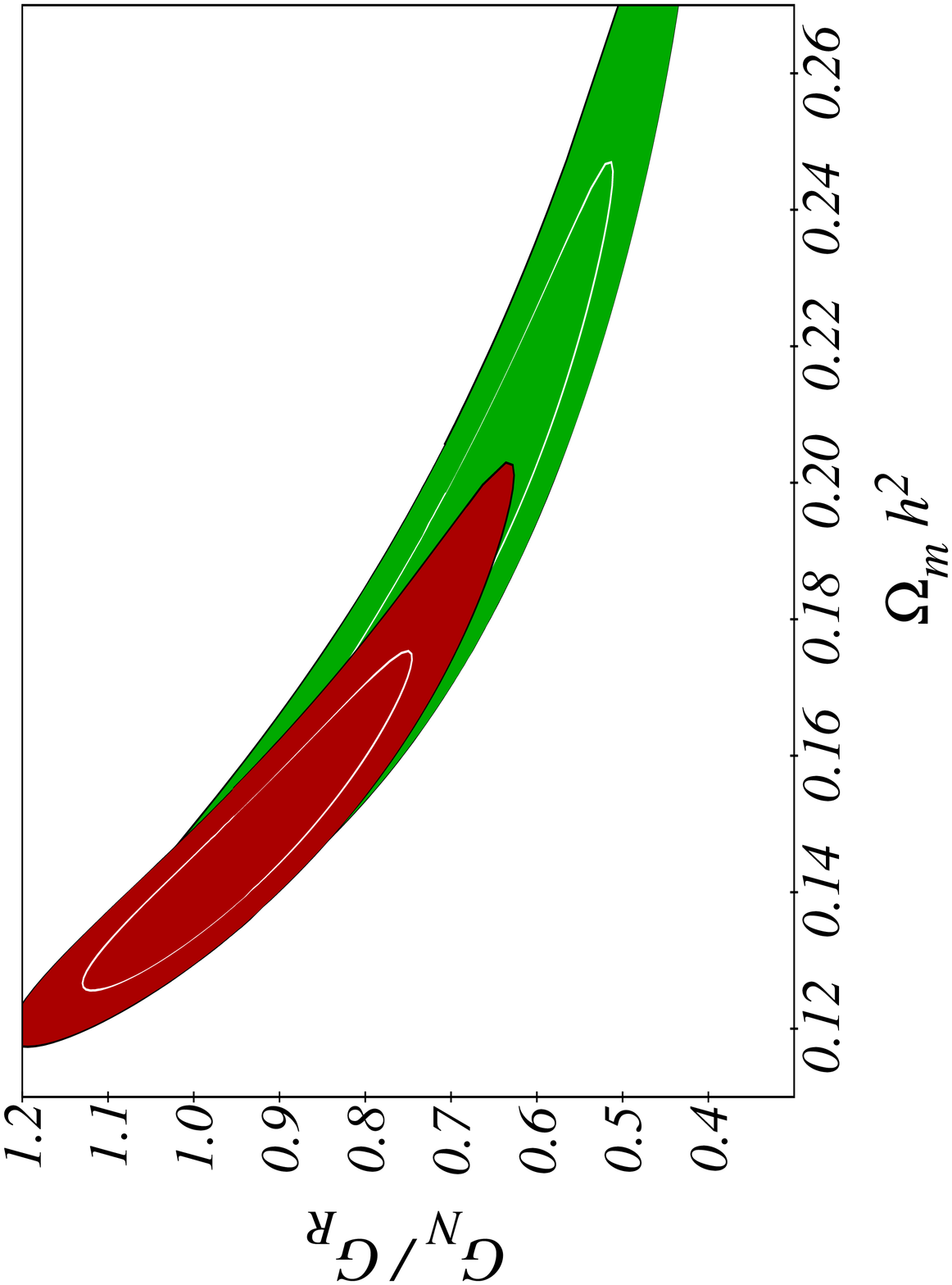}
\includegraphics[bb=0 0 595.3  841.9, angle=-90, width=0.3\textwidth]{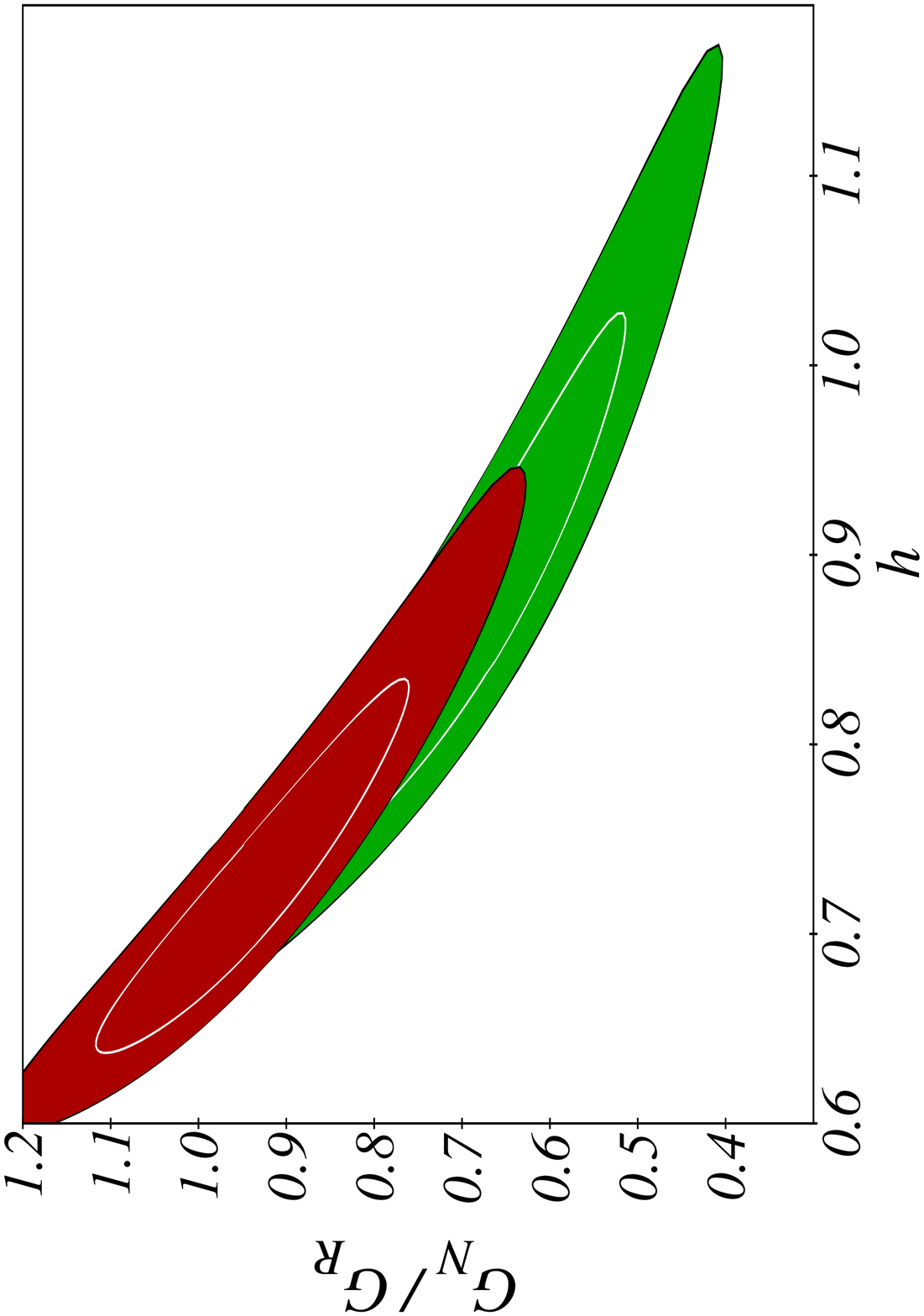}
\includegraphics[bb=0 0 595.3  841.9, angle=-90, width=0.3\textwidth]{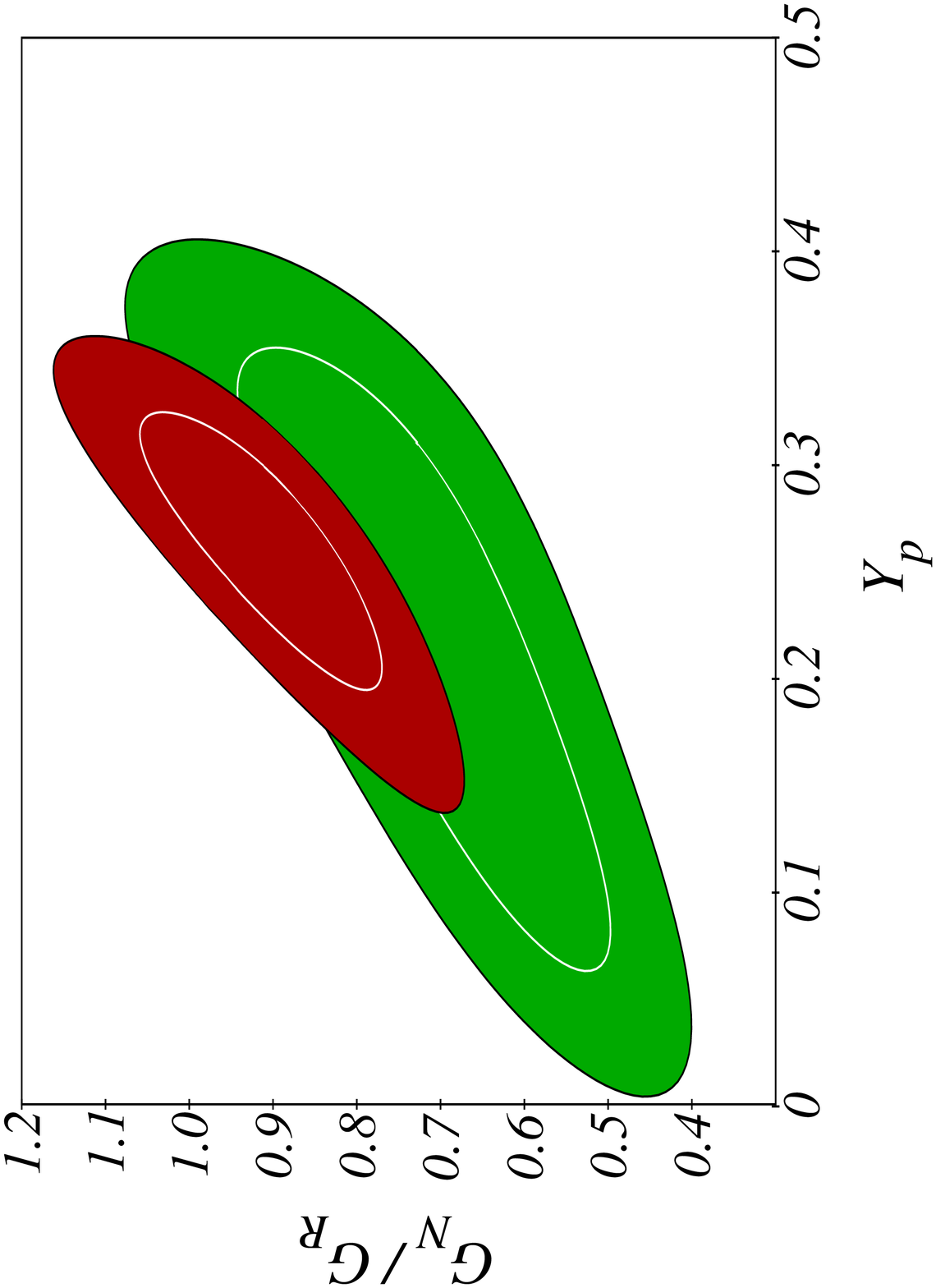}

\captionsetup{margin=10pt,font=small,labelfont=bf}
\caption{Constraints at the $95\%$ confidence level for $G_N/G_R$ from WMAP+ACT+Sne+Ly-$\alpha$ (background, green) and WMAP+ACT+SPT+Sne+Ly-$\alpha$ (front, red).
         The white lines show the $68\%$ confidence levels. Note that the Gravitational Aether prediction is
	$G_N/G_R=0.75$, while in General Relativity $G_R=G_N$.
        }
\label{fig2b}
\end{figure*}
\begin{table}
\begin{tabular}{|l| c |}
\hline
&   ~~~$G_N / G_R$~~~ \\ \hline
WMAP+ACT & $0.73^{+0.31}_{-0.21}$ \\ \hline
WMAP+ACT+SPT & $ 0.88^{+0.17}_{-0.13}$  \\ \hline
WMAP+ACT+Hubble+BAO+Sne & $0.89^{+0.13}_{-0.11} $  \\ \hline
WMAP+ACT+SPT+Hubble+BAO+Sne & $0.94^{+0.10}_{-0.09}$  \\ \hline
WMAP+ACT+Sne+Ly-$\alpha$ (free $Y_p$) & $0.68^{+0.32}_{-0.25} $  \\ \hline
WMAP+ACT+SPT+Sne+Ly-$\alpha$ (free $Y_p$) & $0.90^{+0.27}_{-0.23} $  \\ \hline
\end{tabular}

\captionsetup{margin=10pt,font=small,labelfont=bf}
\caption{Summary of the constraints on $G_N/G_R$ and the associated $95\%$ confidence intervals for different combinations of observational data.}
\label{tab-constraints}
\end{table}

\section{Precision Tests of Gravity}
Gravity on millimeter to solar system scales is well described by General Relativity, which has passed many precision tests on these scales with flying colors (see {\em e.g.,} \cite{Will:2005va} for a review). That is why it is hard to imagine how an order unity change in the theory such as that of Eq. (\ref{1}) can be consistent with these tests, without introducing any fine-tuned parameter. In this section, we argue that nearly all these tests are with gravitational sources that have negligible {\em pressure} or {\em vorticity}, which source deviations from GR predictions in gravitational aether theory.

\subsection{Parametrized Post-Newtonian (PPN) formalism}
In Sec. \ref{cosmology}, we argued that for any perfect fluid with constant equation of state, $w$, the solutions of gravitational aether theory are identical to those of GR with a renormalized gravitational constant $\propto (1+w)$. However, for generic astrophysical applications, $w$ is not constant except for pure radiation, or in the pressureless limit of $w=0$. Focusing on the latter case, and given that pressure is 1st order in post-Newtonian expansion, we can quantify the gravitational aether theory through the Parametrized Post-Newtonian (PPN) formalism.

The Parametrized Post-Newtonian (PPN) formalism is defined in a weak field,
slow motion limit, and describes the next-to-Newtonian order gravitational
effects in terms of a standardized set of potentials and ten parameters.
These PPN parameters will be determined by solving the field equations
(\ref{1}) order-by-order with a perfect fluid source in a standard coordinate gauge.
The conventional introductory details of the formalism will be skipped over
(see \cite{Foster:2005dk} for a more detailed explanation of the procedure and the
general PPN formalism).

To be clear, though, we will assume a nearly globally Minkowskian coordinate
system and basis with respect to which, at zeroth order, the metric is the
Minkowski metric ($g_{\mu\nu}=\eta_{\mu\nu}$) and the fluid four-velocity
$u^{\mu}$ is purely timelike ($u^0=1, u^i=0$). The stress-energy tensor is taken
to have the form $T_{\mu\nu} = (\rho+\rho\Pi+p) u_{\mu}u_{\nu} + pg_{\mu\nu}$
where $u_{\mu}$, $\rho$, $\Pi$ and $p$  are the the unit four-velocity, rest-mass-energy
density, internal energy density, and isotropic pressure of the fluid source, respectively.
The fluid variables are assigned orders of
$\rho \sim \Pi \sim \frac{p}{\rho} \sim u_i^2 \sim1\mathrm{PN}$. In the weak field limit,
the metric can be written as a purturbation of the Minkowski metric:
$g_{\mu\nu}=\eta_{\mu\nu}+h_{\mu\nu}$. The components of the metric perturbations
$h_{\mu\nu}$ with respect to this basis will be assumed to be of orders:
$h_{00}\sim1\mathrm{PN} + 2\mathrm{PN}$, $h_{ij}\sim1\mathrm{PN}$, and $h_{0i}\sim1.5\mathrm{PN}$. This choice
preserve the Newtonian limit while allowing one to determine just the first
post-Newtonian corrections. Furthermore, the aether four-velocity $u'_{\mu}$
will be assumed to be of the same order as that of the matter fluid.

Solving (\ref{3}) to 1PN gives $p'=-\rho/4$, which can be used in (\ref{1})
to solve for $g_{00}$ and $g_{ij}$ to 1PN:
\beq
h_{00}=2U
\label{6}
\eeq
\beq
h_{ij}=2U\delta_{ij},
\label{7}
\eeq where $U$ is the Newtonian potential and the following gauge condition
is imposed: $\partial_j h_{ij} =\frac{1}{2}(\partial_i h_{jj}-\partial_i h_{00})$.
Comparing the continuity equations for matter and aether (i.e. the "time"
component of (\ref{3}) to 1.5 PN), it can be shown that
\beq
u'^i-u^i=t^i,
\eeq where $t^i$ satisfies $\nabla^i(t_i\rho)=0$. This implies that the
rotational component of aether is not fixed by matter within the PN expansion 
formalism. Here we will make the assumption that 
$t^i=0$ so that aether is completely dragged by matter. We will discuss this choice further in Section (\ref{gravitomag}). 

Previously we mentioned that in this case, an exact solution for $u'_{\mu}$ and $p'$ exists when matter has a constant
equation of state. (It is worth noting that in the $t^i = 0$ case, higher PN equations appear to 
imply a nonstandard condition on the pressure $\nabla_a (u^a p)$ = 0, which is satisfied for a 
constant equation of state.) Using this solution and
an additional gauge condition $\partial_ih_{0i}=3\partial_0U$, the field
equations can be solved for $g_{0i}$ and $g_{00}$ to 1.5PN and 2PN,
respectively:
\beq
h_{0i}=-\frac{7}{2}V_i-\frac{1}{2}W_i,
\label{8}
\eeq
\beq
h_{00}=2U-2U^2+4\phi_{1}+4\phi_{2}+2\phi_{3}+6(1+\frac{1}{3})\phi_{4},
\label{9}
\eeq where Appendix \ref{app1} includes the definition for all potentials.
Collecting all the results (\ref{6}), (\ref{7}), (\ref{8}), and
(\ref{9}) indicates that all metric components are as in standard GR,
except for the term in $g_{00}$ with the pressure-dependent potential $\zeta_4$.
Consulting the parametrization rubric indicates that all PPN parameters
have the standard values except $\zeta_4$, which equals
\beq
\zeta_4=\frac{1}{3},
\label{73}
\eeq
which was already pointed out in \cite{Afshordi:2008xu}. Notice that $\zeta_4$, {\em i.e.} the anomalous coupling of gravity to pressure is the only PPN parameter that is not measured experimentally, as one needs to probe the relationship between gravity and pressure of an object with near-relativistic pressures. A notable exception is observation of neutron stars (or their mergers, via gravitational wave observations), which can potentially measure $\zeta_4$, assuming that the uncertainties in nuclear equation of state are under control \cite{Farbod}.

\subsection{Gravitomagnetic Effect}
\label{gravitomag}
In the previous section, we showed that rotation of aether is not
fixed by matter in the non-relativistic regime. We further assumed that
aether rotates with matter. Here we will argue that observational
bounds on the gravitomagnetic effect provide a mild prefernce for this
assumption.

Space-time around a rotating object with a weak
gravitational field, like Earth, can be described in terms of a set of
potentials. With appropriate definitions, these potentials satisfy
equations analogous to Maxwell's equations \cite{Ruggiero:2002hz}.
The gravitomagnetic effect describes the dragging of spacetime
around a rotating object and can be
quantified by a gravitomagnetic field $\vec{B}$ defined as:
\beq
\vec{B}=-4\frac{3\vec{r}(\vec{r}\cdot\vec{S})-\vec{S}r^2}{2r^5},
\eeq
\beq
S^i=2G'\int \epsilon^i_{jk}x'^{j}T_{\rm eff}^{0k} \mathrm{d}^3x'.
\eeq where $\vec{r}$ is the position vector measured from the center of the object,
$\epsilon^i_{jk}$ is the three-dimensional Levi-Civita
tensor, and $T_{\rm eff}^{\mu\nu}$ is the RHS of the field equations (\ref{1}) \cite{Ruggiero:2002hz}.
The gravitomagnetic field causes the precession of the orbital
angular momentum of a free falling test particle. The angular velocity
of this precession is \cite{Ruggiero:2002hz}
\beq
\vec{\Omega}=-\frac{\vec{B}}{2}.
\eeq
If aether is irrotational, $T_{\rm eff}^{0k} =T^{0k}$ to within the accuracy
of linearized theory and since $G'=\frac{4}{3}G_N$, we have:
\beq
\vec{\Omega}_{aether}=\frac{4}{3}\vec{\Omega}_{GR}.
\eeq

Gravity Probe B (GP-B) is an experiment that measures the precession rate $<\Omega>$
of four gyroscopes orbiting the Earth. Recently, GP-B reported a frame-dragging drift rate of
$-37.2\pm7.2$ mas/yr, to be compared with the GR prediction of $-39.2$ mas/yr ('mas' is milliarc-second) \cite{Everitt:2011hp}. Laser ranging to the  LAGEOS and LAGEOS II satellites also provides a measurement of the
frame-dragging effect. The total uncertainty in this case is still being debated; with optimistic estimates of
$10\%-15\%$ ({\it e.g.,} \cite{Ciufolini:2004rq}), and more conservative  estimates as large as $20\%-30\%$ ({\it e.g.,} \cite{Iorio:2008vm}).

Therefore, we conclude that even though perfect co-rotation of aether by matter is preferred by current tests of intrinsic gravitomagnetic effect, an irrotational aether is still consistent with present constraints at $2\sigma$ level.

\section{Conclusions and Discussions}

In the current work, we studied the phenomenological implications of the gravitational aether theory, a modification of Einstein's gravity that solves the old cosmological constant problem at a semi-classical level. We showed that the deviations from General Relativity can only be significant in situations with relativistic pressure, or (potentially) relativistic vorticity. The most prominent prediction of this theory is then that gravity should be 33\% stronger in the cosmological radiation era than GR predictions. We showed that many cosmological observations, including CMB (with the exception of SPT), Ly-$\alpha$ forest, and ${}^{7}\text{Li}$ primordial abundance prefer this prediction, while deuterium may prefer GR values. We then examined the implications for precision tests of gravity using the PPN formalism, and showed that the only PPN parameter that deviates from its GR value is $\zeta_4$, the anomalous coupling to pressure, that has never been tested experimentally. Finally, we argued that current tests of Earth's gravitomagnetic effect mildly prefer a co-rotation of aether with matter, although they are consistent with an irrotational aether at $2\sigma$ level.

In our opinion, the fact that gravitational aether has {\it the same number of free parameters as GR}, and is yet (to our knowledge) consistent with all cosmological and precision tests of gravity at
2$\sigma$
level, indicates that this theory could be a strong contender for Einstein's theory of gravity.

Future observations are expected to sharpen these distinctions. In particular, the most clear test will come from the Planck CMB anisotropy power spectrum that is expected to be released in early 2013. Judging by the predictions for constraints on the effective number of neutrinos, Planck should be able to distinguish GR and Aether at close to 10$\sigma$ level \cite{Hamann:2010bk}. 

Another interesting implication of this theory is for the cosmic baryon fraction. As we increase the gravity due to radiation (or effective number of neutrinos), we need to increase the dark matter density to keep the redshift of equality constant, since it is well constrained by CMB power spectrum (see {\it e.g.,} ~\cite{Komatsu:2010fb}). This implies that the total matter density should be bigger by a factor of $4/3$ (Fig. \ref{fig2a}). Given that baryon density is insensitive to this change, the cosmic baryon fraction will decrease by a factor of $3/4$, i.e. from 17\% \cite{Komatsu:2010fb} to 13\%. This could potentially resolve the missing baryon problem in galaxy clusters \cite{Afshordi:2006pc}, as well as the deficit  in observed Sunyaev-Zel'dovich power spectra, in comparison with theoretical predictions \cite{Lueker:2009rx,Dunkley:2010ge}.


\acknowledgments

We would like to thank Tom Giblin, Ted Jacobson, Justin Khoury, Maxim Pospelov, Josef Pradler, Bob
Scherrer, and Kris Sigurdson for useful discussions and comments throughout the
course of this project.  GR thanks the Perimeter Institute for
hospitality. SA and NA are supported by the University of Waterloo and the
Perimeter Institute for Theoretical Physics. Research at Perimeter
Institute is supported by the Government of Canada through Industry
Canada and by the Province of Ontario through the Ministry of Research
\& Innovation. K.K. was partly supported by the Grant-in-Aid for the
Ministry of Education, Culture, Sports, Science, and Technology,
Government of Japan, No. 18071001, No.  22244030, No. 21111006, and
No. 23540327, and by the Center for the Promotion of Integrated
Sciences (CPIS) of Sokendai.

\appendix
\section{Aether perturbations through equality}
\label{app2}
Here we present a consistent treatment of cosmological scalar perturbation theory for Gravitational Aether (GA).
As we argued in Section \ref{cosmology}, when matter is approximated by a perfect fluid with density $\rho$, pressure $p=w\rho$ ($w$ constant), and
four velocity $u^{\mu}=\frac{dx^{\mu}}{\sqrt{-ds^2}}$ (i.e.
$T_{\mu\nu}=(1+w)\rho u_{\mu}u_{\nu}+w\rho g_{\mu\nu}$),
$u'_{\mu}=u_{\mu}$ and $p'=\frac{(1+w)(3w-1)}{4}\rho$ solve \eqref{aether_cont} and \eqref{aether_euler} and the GA field equation
\eqref{1} becomes
\beq
(8\pi )^{-1}G_{\mu\nu}=G_N(1+w)T_{\mu\nu}.
\eeq In cosmology, therefore, if the constituents of the universe are matter and radiation \textit{and they are separately conserved},
the GA field equations become
\beq
(8\pi )^{-1}G_{\mu\nu}=G_NT^m_{\mu\nu}+\frac{4}{3}G_NT^r_{\mu\nu},
\eeq where \textit{m} and \textit{r} stand for matter and radiation respectively. This approximation, of course, is false when inhomogeneities are
considered since baryons and photons interact strongly. Therefore, we shall perturb about this exact solution and do a consistent treatment of cosmological scalar perturbation
theory.

In what follows, \textit{b}, \textit{dm}, \textit{m}, and \textit{r} stand for baryon, dark matter, matter, and radiation respectively, and all barred quantities are unperturbed.
Following \cite{ma-1995-455},  we will use the Conformal Newtonian Gauge:
\beq
ds^2=a^2(\tau)\{-[1+2\psi(\tau,\vec{x})]d\tau^2+[1-2\phi(\tau,\vec{x})]dx^idx_i\}.
\eeq
To linear order in perturbation theory, the matter energy-momentum tensor takes the form
\bea
T^0_{\phantom{0}0}&=&-(\bar{\rho}+\delta\rho)\\
T^0_{\phantom{0}i}&=&(\bar{\rho}+\bar{p})\frac{\delta u_i}{a}\\
T^i_{\phantom{i}j}&=&(\bar{p}+\delta p)\delta_{ij}+\Sigma^i_{\phantom{i}j},
\eea
where $\Sigma^i_{\phantom{i}j}$ is the traceless anisotropic shear stress perturbation and
\beq
\delta\rho=\rho-\bar{\rho}; \delta p=p_{i}-\bar{p};\delta u^{i}_{\mu}= u^{i}_{\mu}-\bar{u}_{\mu},
\eeq
where $i=\{dm,b,r\}$.
In our coordinate system $\bar{u}^0=\frac{1}{a}$, $\bar{u}_0=-a$, and $\bar{u}_i=\bar{u}^i=0$.
The Aether pressure and four-velocity perturbations are defined as follows:
\beq
p'=-\frac{\rho_m}{4}+\delta p', u'_{\mu}=u^{dm}_{\mu}+\delta u_{\mu}.
\eeq
Dark matter only interacts gravitationally and is separately conserved. We assume that there is negligible energy transfer between baryons and relativistic particles (i.e. $\nabla^{\mu}(\rho_bu^b_{\mu})=0$). Then, to first order in perturbation theory \eqref{aether_cont} and \eqref{aether_euler} give:
\bea
3\frac{\dot{a}}{a^2}\delta p'&=&\frac{\bar{\rho}_m}{4}\partial_i(\delta u^i+\frac{\bar{\rho}_b}{\bar{\rho}_m}\delta w^i)\label{AP2}\\
\partial_i\delta p'&=&\frac{a\bar{\rho}_m}{4}(\delta\dot{u}^i+2\frac{\dot{a}}{a}\delta u^i)
\label{AP1},
\eea
 where $\delta w^i=\delta u_{dm}^i-\delta u_{b}^i=a^{-2}(\delta u^{dm}_i-\delta u^{b}_i)$ and $\delta u^i=a^{-2}\delta u_i$.  Taking the comoving divergence of \eqref{AP1} and applying the comoving Laplacian to \eqref{AP2}, we can eliminate $\delta p'$ and get an equation for $\Omega\equiv\partial_i\delta u^i$:
\beq
3\frac{\dot{a}}{a^3}\partial_{\tau}(a^2\Omega)-\nabla^2\Omega=\frac{\bar{\rho}_b}{a\bar{\rho}_m}\nabla^2(\dot{\delta}_b-\dot{\delta}_{dm}),
\label{AP3}
\eeq
where $\delta_{dm}=\frac{\delta\rho_{dm}}{\bar{\rho}_{dm}}$, $\delta_{b}=\frac{\delta\rho_{b}}{\bar{\rho}_b}$,
 and we have used the fact that $\partial_i\delta w^i=\frac{1}{a}(\dot{\delta}_b-\dot{\delta}_{dm})$. In Fourier space, this equation can be numerically integrated for modes of different wavelength, given the equations that govern $\delta_{dm}$ and $\delta_{b}$. Once $\Omega$ is known, \eqref{AP2} can be used to find $\delta p'$.
In the Conformal Newtonian Gauge, only scalar perturbations are treated and we can ignore the rotational part of $\delta u^i$. This can also be physically motivated: let $\delta u^i=\partial_iu_S+\partial\delta u^i_V$ where $\partial_i\delta u^i_V=0$. Taking the curl of \eqref{AP1}, it follows that $\nabla\times\delta\vec{u}_V\propto\frac{1}{a^2}$.
As a result, the rotational part of the Aether fluid decays and it doesn't play a major role in cosmology.
As a result, given $\Omega$ we can find $\delta u^i$ in Fourier space ($\partial_j\to ik_j$):
\beq
\delta u^j=-i\frac{k_j}{k^2}\Omega,
\eeq
 where $k^2=\delta_{ij}k_ik_j$. Similarly,
\beq
\delta w^j=\delta u_{dm}^j-\delta u_{b}^j=i\frac{k_j}{ak^2}(\dot{\delta}_{dm}-\dot{\delta}_b).
\eeq
To first order in perturbation theory, the GA field equations now take the form
\beq
(8\pi G_N)^{-1}G_{\mu\nu}=T^m_{\mu\nu}+\frac{4}{3}T^r_{\mu\nu}+\epsilon_{\mu\nu}
\eeq with $\epsilon_{00}=0$, $\epsilon_{0i}=\frac{a\bar{\rho}_m}{3}\big(\delta u_i+\frac{\bar{\rho}_b}{\bar{\rho}_m}\delta w_i\big)$, and $\epsilon_{ij}=\frac{4}{3}a^2\delta p' \bar{g}_{ij}$.
\\\\
Having both the left and right hand sides of this equation, we can now solve for the scalar perturbations.
However, this does not provide an obvious way of checking the prediction of this theory, namely $\frac{G_R}{G_N}=\frac{4}{3}$.
This can be easily accommodated for by having field equations that contain $G_R$ as a constant, and reduce to General Relativity and GA for $G_R=G_N$
and $G_R=\frac{4}{3}G_N$ respectively.
Consider the field equations (which we will refer to as the Generalized Gravitational Aether (GGA) field equations)
\bea
(8\pi)^{-1}G_{\mu\nu}[g_{\mu\nu}]&=&G_RT_{\mu\nu}+(G_N-G_R)T^{\alpha}_{\alpha}g_{\mu\nu}+\tilde{T}_{\mu\nu} \notag\\
\tilde{T}_{\mu\nu}&=&\tilde{p}(\tilde{u}_{\mu}\tilde{u}_{\nu}+g_{\mu\nu}). \label{GFE}
\eea
Conservation of $G_{\mu\nu}$ and $T_{\mu\nu}$ implies
\beq
\nabla^{\mu}\tilde{T}_{\mu\nu}=(G_R-G_N)\nabla_{\nu}T.
\label{GCC}
\eeq Defining $p'=\frac{\tilde{p}}{4(G_R-G_N)}$ and making the obvious identification $\tilde{u}_{\mu}=u'_{\mu}$,  we see that this equation becomes exactly \eqref{3}. Therefore, all of our solutions before can be used here after a trivial rescaling of the pressure.
For example, if $T_{\mu\nu}$ is a perfect fluid with equation of state $w$, exact solutions are obtained by $\tilde{u}_{\mu}=u_{\mu}$ and $\tilde{p}=(G_R-G_N)(1+w)(3w-1)\rho$, which again just renormalizes Newton's constant:
\beq
G_N\longrightarrow G_{eff}(w)=G_N\big\{3w\frac{G_R}{G_N}+(1-3w)\big\}.
\eeq
Note that $G_{eff}(w=0)=G_N$ and $G_{eff}(w=1/3)=G_R$. Again, if matter and radiation are separately conserved in a cosmological setting, \eqref{GFE} becomes
\beq
(8\pi )^{-1}G_{\mu\nu}=G_NT^m_{\mu\nu}+G_RT^r_{\mu\nu}.
\eeq
More importantly, when $G_R=G_N$, these field equations reduce to those of General Relativity (GR) (this is true in the cosmological case because $\nabla_{\mu}\tilde{u}^{\mu}\neq0$, which means that the conservation of Aether implies that its pressure vanishes identically). Also when $G_R=\frac{4}{3}G_N$, the GAA field equations reduce to those of GA, with the appropriate rescaling $T'_{\mu\nu}=\frac{3}{4G_N}\tilde{T}_{\mu\nu}$.
Therefore, fitting this theory to data, we will be able to make a likelihood plot of $\frac{G_R}{G_N}$ and see how far away the best fit is from the GA and GR predictions.
\\\\
Because of the similarity of the underlying equations, the linear perturbation theory of the GGA field equations is very close to those of GA, which we already described. We treat all matter perturbations as before and perturb $\tilde{T}_{\mu\nu}$ similarly:
\beq
\tilde{p}=(G_N-G_R)\rho_m+\delta\tilde{p},\tilde{u}_{\mu}=u^{dm}_{\mu}+\delta u_{\mu}.
\eeq The equations of interest are (in Fourier space):
\beq
3H\partial_{\tau}(a^2\Omega)+a(\tau)k^2\Omega=k^2\frac{\bar{\rho}_{b_0}}{\bar{\rho}_{m_0}}(\dot{\delta}_{dm}-\dot{\delta}_b) \label{Omega_Eq}
\eeq
\begin{align}
\delta \tilde{p}&=\frac{(G_R-G_N)\bar{\rho}_m}{3H}\big[\Omega+\frac{\bar{\rho}_{b_0}}{a\bar{\rho}_{m_0}}(\dot{\delta}_b-\dot{\delta}_{dm})\big]\label{deltap}\\
\delta \tilde{u}^j&=-i\frac{k_j}{k^2}\Omega, \label{deltau}
\end{align}
where $H=\frac{\dot{a}}{a^2}$ and we have recognized that $\frac{\bar{\rho}_b}{\bar{\rho}_m}=\frac{\bar{\rho}_{b_0}}{\bar{\rho}_{m_0}}$ is fixed by the values at the present time.
Once \eqref{Omega_Eq} is solved for $\Omega$, $\delta\tilde{p}$ and  $\delta \tilde{u}^j$ are determined by \eqref{deltap} and \eqref{deltau}, respectively.
At long last, to linear order in perturbation theory, the GAA field equations read
\beq
(8\pi)^{-1}G_{\mu\nu}=G_NT^m_{\mu\nu}+G_RT^r_{\mu\nu}+\tilde{\epsilon}_{\mu\nu}, \label{FinalFE}
\eeq
where
\begin{align}
\tilde{\epsilon}_{00}&=0\\
\tilde{\epsilon}_{0j}&=i\frac{k_j}{k^2}(G_N-G_R)(a^3\bar{\rho}_m)\big[\Omega+\frac{\bar{\rho}_{b_0}}{a\bar{\rho}_{m_0}}(\dot{\delta}_b-\dot{\delta}_{dm})\big]\\
\tilde{\epsilon}_{ij}&=(G_R-G_N)\frac{\bar{\rho}_ma^2}{3H}\big[\Omega+\frac{\bar{\rho}_{b_0}}{a\bar{\rho}_{m_0}}(\dot{\delta}_b-\dot{\delta}_{dm})\big]\delta_{ij}.
\end{align}
Having both the left and right hand sides of \eqref{FinalFE}, the scalar perturbations can now be consistently solved for.
\\
\section{PPN notations}
\label{app1}
The metric components are in terms of particular potential functions,
thus defining the PPN parameters:
\bea
g_{00} &=& -1+2U-2\beta U^2 -2\xi\phi_W+(2\gamma+2+\alpha_3+\zeta_1-2\xi)\phi_1 \nonumber \\
             && +\:2(3\gamma-2\beta+1+\zeta_2+\xi)\phi_2+2(1+\zeta_3)\phi_3\nonumber \\
             &&+\:2(3\gamma+3\zeta_4-2\xi)\phi_4-(\zeta_1-2\xi)A
\\
g_{ij} &=& (1+2\gamma U)\delta_{ij}
\\
g_{0i} &=& -\frac{1}{2}(4\gamma+3+\alpha_1-\alpha_2+\zeta_1-2\xi)V_i\nonumber \\
        && -\:\frac{1}{2}(1+\alpha_2-\zeta_1+2\xi)W_i
\eea
The potentials are all of the form
\beq
F(x)=G_N \int \mathrm{d}^3y\, \frac{\rho(y)f}{|x-y|}
\eeq and the correspondences $F:f$ are given by
\bea
U&:&1 \qquad \phi_1:u_iu_j \qquad  \phi_2:U \qquad  \phi_3:\Pi \qquad  \phi_4:p/\rho \notag\\
\phi_W&:&\int \mathrm{d}^3z\,\rho(z) \frac{(x-y)_j}{|x-y|^2}\Big[\frac{(y-z)_j}{|x-z|}- \frac{(x-z)_j}{|y-z|}\Big]\notag\\
A&:&\frac{(v_i(x-y)_i)^2}{|x-y|^2}\notag \\
V_i&:&u^i \qquad W_i:\frac{u_j(x_j-y_j)(x^i-y^i)}{|x-y|^2}.
\eea

\bibliography{AetherObservations_6,KazRef-2,cmbrefs}

\end{document}